# INFRARED SPECTRA OF METHYL-, AND NITROGEN-MODIFIED VOID CORONENE; MODELING A CARRIER OF INTERSTELLAR POLYCYCLIC AROMATIC HYDROCARBON

NORIO OTA
Graduate School of Pure and Applied Sciences, University of Tsukuba,
1-1-1 Tenoudai Tsukuba-city 305-8571, Japan; n-otajitaku@nifty.com

Void induced coronene $C_{23}H_{12}^{++}$ was suggested to be a possible carrier of the astronomically observed polycyclic aromatic hydrocarbon (PAH), which shows unique molecular structure with carbon two pentagons connected with five hexagons. Well observed astronomical infrared spectrum from 3-15μm could be almost reproduced based on density functional theory. However, there remain several discrepancies with observed spectra, especially on 11-15μm band weaker intensity. Observed 11.2μm intensity is comparable to 7.6-7.8μm one. Methyl-modified molecule $C_{24}H_{14}^{++}$ revealed that calculated peak height of 11.4μm show fairly large intensity up to 70-90% compared with that of 7.6-7.8μm band. Also, nitrogen atom was substituted to peripheral C-H site of void coronene to be $C_{22}H_{11}N_1^{++}$. Pentagon site substituted case show 60% peak height. This molecule also reproduced well 12-15μm peak position and relative intensity. Vibration mode analysis demonstrated that 11.3μm mode comes from C-H out of plane bending. Heavy nitrogen plays as like an anchor role for molecule vibration.

## 1, INTRODUCTION

Interstellar organic material is a key issue to find a missing link between basic astronomical carbon material and biological components on the earth. It was so called "organic chemical evolution in space". Interstellar dust show mid-infrared emission from 3 to 20μm. Discrete emission features at 3.3, 6.2, 7.6, 7.8, 8.6, 11.2, and 12.7μm are ubiquitous peaks observed at many astronomical objects (Ricca et al. 2012; Geballeet al. 1989; Verstraete et al. 1996; Moutou et al. 1999; Meeus et al. 2001; Peeters et al. 2002; Regan et al.2004; Engelbracht et al. 2006; Armus et al. 2007; Smith et al. 2007; Sellgren et al. 2007). Current understanding is that these astronomical spectra come from the vibrational modes of polycyclic aromatic hydrocarbon (PAH) molecules. Concerning PAH spectra, there are many experimental (Szczepanski & Vala 1993a; Schlemmer et al. 1994; Moutou et al. 1996; Cook et al. 1998; Piest et al. 1999; Hudgins& Allamandola 1999a, 1999b; Oomens et al. 2001, 2003, 2011;Kim et al. 2001) and density functional theory (DFT) based theoretical analysis (de Frees et al. 1993; Langhoff 1996;Malloci et al. 2007; Pathak & Rastogi 2007; Bauschlicher et al.2008, 2009; Ricca et al. 2010, 2011b, 2012).

The current central concept to understand the observed astronomical spectra is the decomposition method from the data base of many PAHs experimental and theoretical analysis (Boersma et al. 2013, 2014). Recently, Tielens (Tielens 2013) discussed that void induced graphene like PAH's may be one of candidates of emission source. In a previous paper (Ota 2014b), as an example of typical PAH, void induced coronene $C_{23}H_{12}^{++}$ was studied using DFT calculation method. Remarkable result was a similar IR spectrum tendency with astronomically well observed one. Detailed vibrational mode analysis was also opened (Ota, 2015). Such results suggested the existence of a limited set of survived molecules in a harsh environment in space. However, there remain several discrepancies with observed spectra, especially on 11-15μm band weaker intensity. Observed 11.2μm intensity is comparable or larger than 7.6-7.8μm one. The aim of this paper is to modify void coronene in order to increase 11-15μm intensity. Here, methyl ($CH_3$) and Nitrogen were tested to modify void coronene. Fairly large increase of IR intensity was obtained. Vibrational mode analysis was also done.

## 2, CALCULATION METHOD

We have to obtain total energy, optimized atom configuration, and infrared vibrational mode frequency and strength depend on a given initial atomic configuration, charge and spin state Sz. Density functional theory (DFT) with unrestricted B3LYP functional (Becke 1993) was applied utilizing Gaussian09 package (Frisch et al. 2009, 1984) employing an atomic orbital 6-31G basis set. The first step calculation is to obtain the self-consistent energy, optimized atomic configuration and spin density. Required convergence on the root mean square density matrix was less than $10^{-8}$ within 128 cycles. Based on such optimized results, harmonic vibrational frequency and strength was calculated. Vibration strength is obtained as molar absorption coefficient ε (km/mol). Comparing DFT harmonic wavenumber $N_{DFT}$ ($cm^{-1}$) with experimental data, a single scale factor 0.965 was used. Based on experimental data (Hudgins et al. 1998, Ono 2010), suitable scale factor was obtained as figured in Appendix 1. It is 0.7% larger than often used factor 0.958 (Ricca et al. 2012). For the anharmonic correction, a redshift of 15cm-1 was applied (Ricca et al. 2012).

Corrected wave number N is obtained simply by $N(cm^{-1}) = N_{DFT}(cm^{-1}) \times 0.965 - 15 (cm^{-1})$.
Also, wavelength λ is obtained by $\lambda(\mu m) = 10000/N(cm^{-1})$.

## 3, MOLECULAR STRUCTURE

Molecular structures were illustrated in Figure 1. Pure coronene $C_{24}H_{12}$ as (A) is non-magnetic (Sz=0/2) having D6h point group symmetry. Carbon single void was created at marked site by red circle. There are six unpaired electrons, which means multiple spin state capability of Sz=6/2, 4/2, 2/2 and 0/2. Most stable one was singlet state Sz=0/2 (Ota 2015). Such a void brings quantum mechanical distortion by the Jahn-Teller effect (Ota; 2011, 2014a, 2015). It is amazing that there cause bond-bond reconstruction. We can see carbon two pentagons connected with five hexagons as illustrated in (B) $C_{23}H_{12}^{++}$ with point group symmetry of C2. Figures show tilt view of molecule image. Modifying peripheral part by $CH_3$- additive, structure changes like (C) $C_{24}H_{14}^{++}$ ($C_{23}H_{11}$-$CH_3^{++}$). Also, substitution by nitrogen atom was supposed as like (D) $C_{22}H_{11}N_1^{++}$, where nitrogen atom is blue ball.

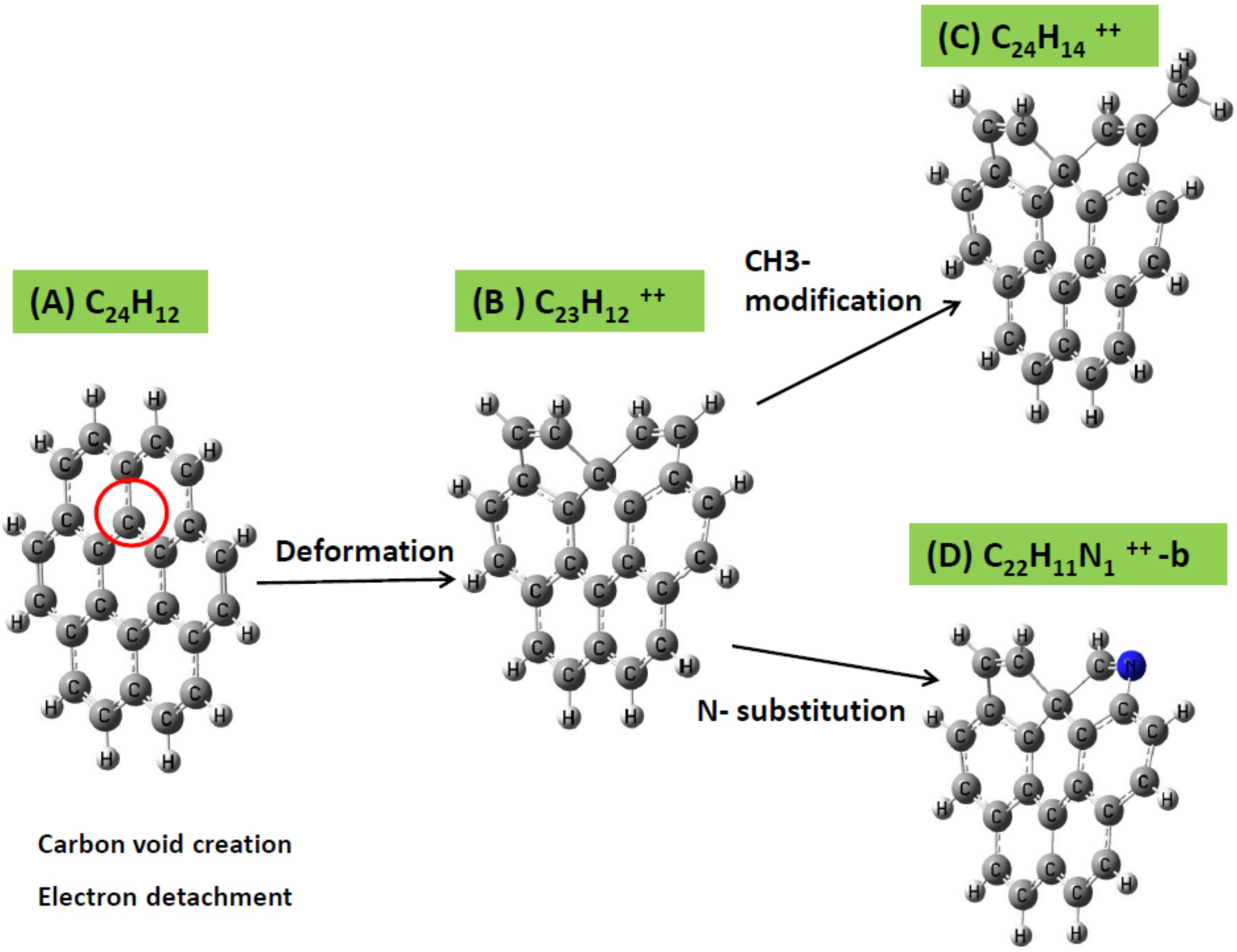

**Figure 1**. Tilt view of typical molecules. By creation of single void and two electron detachment, pure coronene (A) $C_{24}H_{12}$ was deformed to (B) $C_{23}H_{12}^{++}$. Peripheral carbon site is modified by methyl (CH3-) as shown in (C). Also, peripheral carbon is substituted by nitrogen atom as like (D).

## 4, MID-IR SPECTRA

In Figure 2, on a top box, observed spectra from four sources (Boersma et al 2009) were illustrated to compare calculated results. Harmonic wavelength and intensity of molecule (B), (C), and (D) were shown by red diamonds. Vertical value show integrated absorption coefficient ε (km/mol) for every calculated harmonic frequency. Blue curve shows Lorentzian type distribution with the full width at half-maximum (FWHM) of $8cm^{-1}$ based on the harmonic intensity. Vertical value of blue curve is relative and not exactly shown to avoid complexity. Concerning original void induced coronene (B) $C_{23}H_{12}^{++}$, spectrum behavior from 3-15μm was almost similar with observed one. However, there remain several discrepancies with observed spectra, especially on 11-15μm band weaker intensity. Observed 11.2μm intensity is comparable or larger than 7.6-7.8μm one. In $CH_3$-modified case (C), we can see an improved spectrum especially in a region from 11 to 15μm. Calculated peak height of 11.4μm show fairly large intensity up to 70-90% compared with that of 7.6-7.8μm band. Extra influences were observed as a disappearance of 8.6μm band, and extra band at 3.4μm accompanied with 3.2μm. We need careful check with precise observation (Chiar et al. 2000, Steglich et al. 2013). In nitrogen-substituted case (D) $C_{22}H_{11}N_1^{++}$, we could also see an improvement of 11.3μm peak. Pentagon site carbon substituted case ($C_{22}H_{11}N_1^{++}$ -b) show 60% peak height compared with 7.6μm peak. This molecule also reproduced well 12-15μm peak position and relative intensity. We can see large 8.6μm peak and 3.2μm single peak, that is, there are little bad influence compared with case (C).

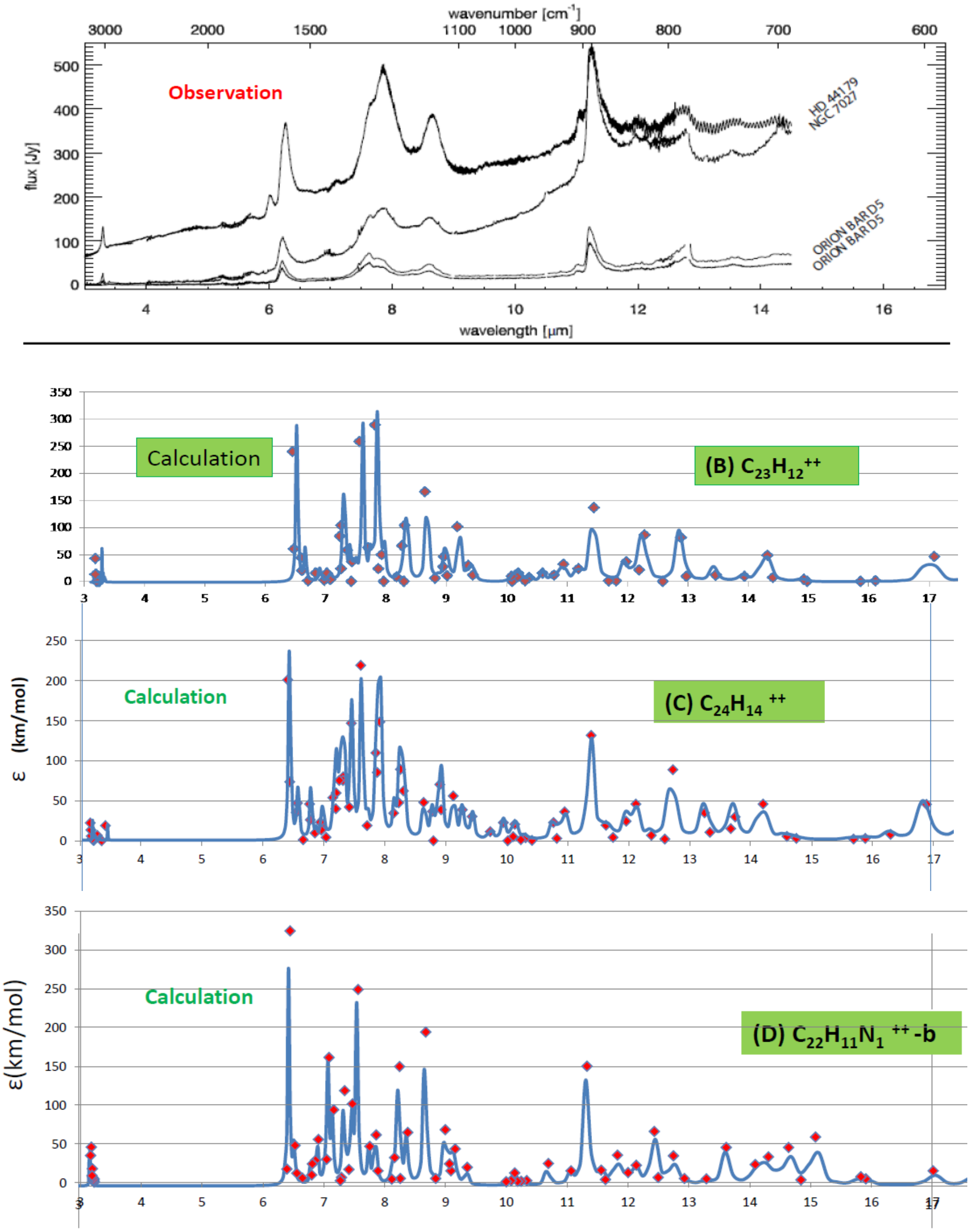

**Figure 2.** Comparison between the observed four sources flux (Boersma et al. 2009) and the calculated infrared spectra of typical void induced and modified coronene family.

## 5, NITROGEN SUBSTITUTED VOID CORONENE

Nitrogen substitution to different carbon position was compared. Typical four position were illustrated on top of Figure 3, that is, pentagon part of -a and -b, also hexagonal part of -c and -d. In case of ($C_{22}H_{11}N_1^{++}$ -a), the increase of 9.3μm peak was recognized, but there appear extra peak at 7.2, 9.0μm. In hexagonal carbon site substituted case ($C_{22}H_{10}N_1^{++}$ -c), only three main peaks were observed at 6.3, 7.5, and 11.5μm, which may not simulate the well observed spectrum. Hexagonal C-H site substituted case was shown in ($C_{22}H_{11}N_1^{++}$ -d), where there are complex peaks from 7-13μm which show no good coincidence with observation. Again, ($C_{22}H_{11}N_1^{++}$ -b) show fairly large 11.3μm peak and reproduced well 12-15μm behavior, large 8.6μm peak and 3.2μm single peak. Comparing these four candidates, case ($C_{22}H_{11}N_1^{++}$ -b) was a better one. Through these examples, we could recognize delicate influence of substituted site on PAH.

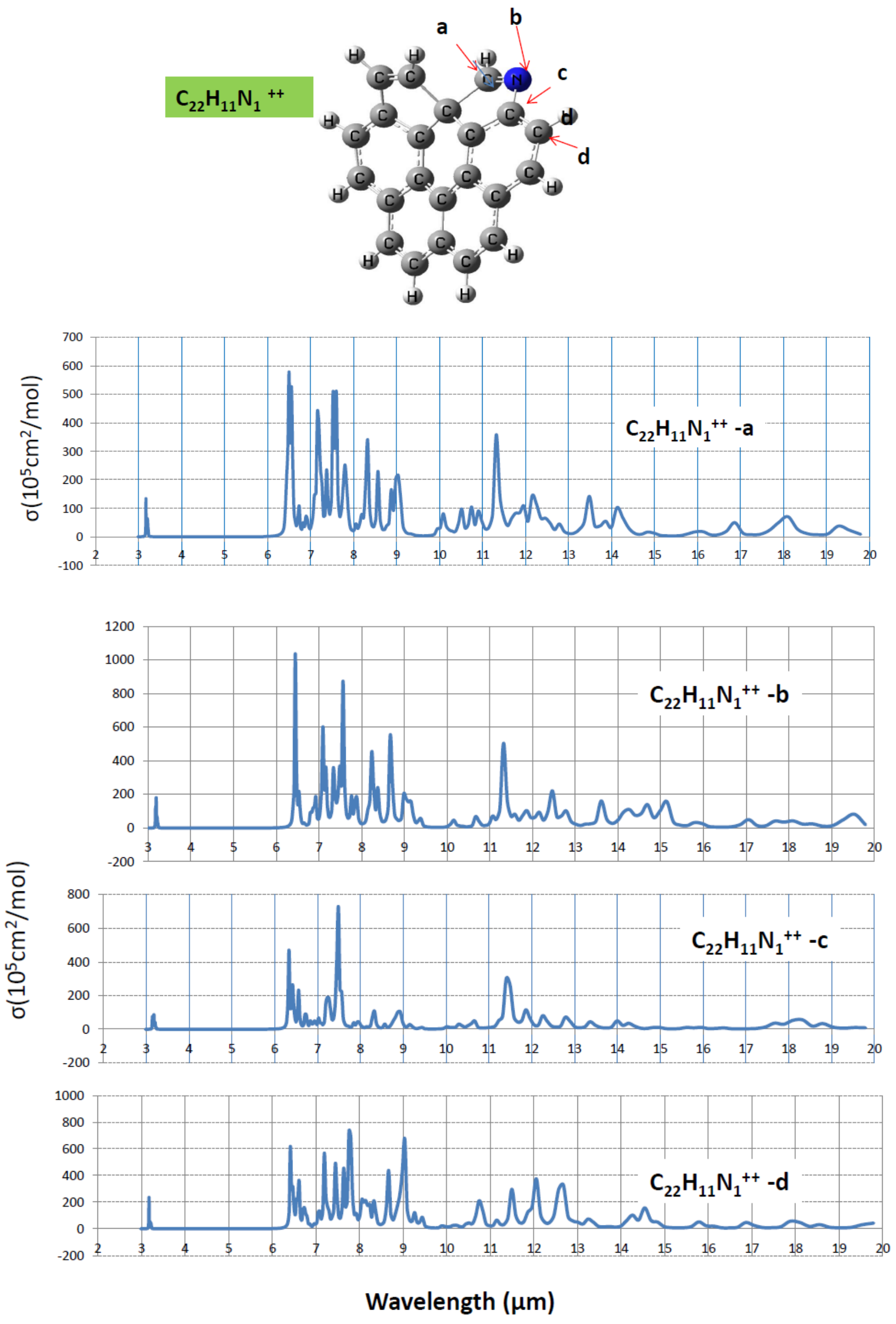

**Figure 3.** Infrared spectra of nitrogen substituted void coronene. Depending on substitute site (a, b, c and d), detailed behavior changes. Most promising candidate is site-b substituted one.

## 6, VIBRATIONAL MODE ANALYSIS OF ($C_{22}H_{11}N_1^{++}$ -b)

All of harmonic frequencies and IR intensities of a promising molecule ($C_{22}H_{11}N_1^{++}$ -b) are noted in Appendix 2. Among them, strong intensity major vibrational modes were analyzed in Table 1. Features are,
(1)3.19μm: C-H stretching mode at nitrogen substituted pentagon site.
(2)6.45μm: C-C stretching at all hexagon sites.
(3)7.1-7.6μm: C-C stretching and C-H in-plane motion
(4)8.7μm: C-H in-plane motion
(4)11.3μm: C-H out-of-plane bending
(5)12.4-15.1μm: C-C stretching
It should be noted that substituted nitrogen is heavy than carbon and play an anchor like role for molecular vibration. These vibration mode features are almost similar to common PAH mode analysis on a review paper (see page 1028, Tielen; 2013)

**Table 1.** Vibrational mode analysis of $C_{22}H_{11}N_1^{++}$ -b

| | Calculation | | | Vibrational mode | Well observed peak |
|---|---|---|---|---|---|
| mode | Ndft(cm-1) | λ(μm) correct | ε(km/mol) | | (μm) |
| 28 | 702.4 | 15.09 | 58.6 | C-C stretching | |
| 30 | 723.2 | 14.64 | 45.3 | C-C streching | 14.3 |
| 33 | 776.4 | 13.62 | 45.2 | C-C streching | 13.5 |
| 38 | 848.7 | 12.44 | 65.6 | C-C streching | 12.7 |
| 44 | 930.2 | 11.33 | 149.9 | C-H out-of-plane bending | 11.2 |
| 59 | 1209.1 | 8.68 | 194.1 | C-H in-plane motion | 8.6 |
| 62 | 1272.0 | 8.25 | 149.6 | C-H in-plane motion | |
| 68 | 1385.4 | 7.56 | 248.7 | C-H in-plane motion, C-C stretching | 7.6 |
| 75 | 1476.6 | 7.09 | 161.2 | C-C stretching | |
| 84 | 1622.7 | 6.45 | 324.3 | C-C stretching | 6.2 |
| 95 | 3268.8 | 3.19 | 45.5 | C-H streching at pentagon | 3.3 |

## 7, CONCLUSION

Modeling a promising carrier of the astronomically observed polycyclic aromatic hydrocarbon (PAH), void induced coronene $C_{23}H_{12}^{++}$ was modified by methylene and nitrogen.

(1) Density functional theory based harmonic frequency analysis was done using a scaling factor of 0.965 to compare experiments.
(2) Well observed astronomical infrared spectrum from 3-15μm could be almost reproduced by $C_{23}H_{12}^{++}$. However, there remain several discrepancies with observed spectra, especially on 11-15μm band weaker intensity. Observed 11.2μm intensity is comparable with 7.6- 7.8μm one.
(3) Methyl-modified molecule ($C_{24}H_{14}^{++}$) revealed that calculated peak height of 11.4μm show fairly large intensity up to 70-90% compared with that of 7.6-7.8μm band.
(4) Nitrogen atom was substituted to peripheral C-H site of void coronene to be $C_{22}H_{11}N_1^{++}$. Pentagon site substituted case show 60% peak height compared with 7.6μm peak., which also reproduced well 12-15μm peak position and relative intensity.
(5) Vibrational mode analysis demonstrated that 11.3μm mode comes from C-H out of plane bending. Heavy nitrogen site acts as like an anchor of molecule vibration.

ACKNOWLEDGEMENT

I would like to say great thanks to Dr. Christiaan Boersma, NASA Ames Research Center, to permit me to refer a figure (Boersma et al. 2009), also thanks his kind directions what information astronomer needs for detailed analysis.

APPENDIX

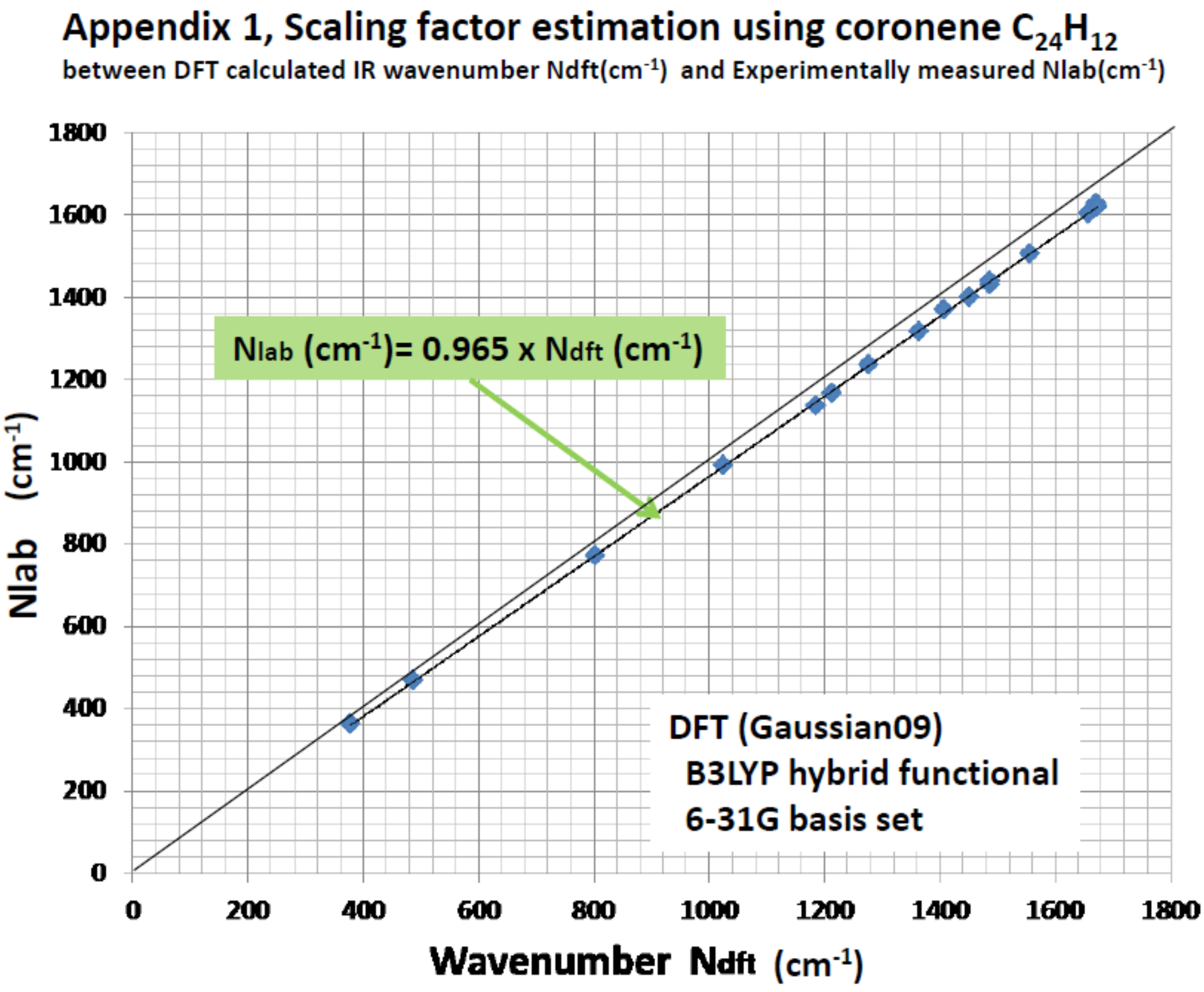

## Appendix 2, $C_{22}H_{11}N_1^{++}$ -b, Harmonic modes

| mode | Ndft(cm-1) | λ(μm)correct | ε(km/mol) |
|---|---|---|---|
| 1 | 97.0449 | 127.1482839 | 2.8572 |
| 2 | 103.4768 | 117.8479383 | 0.6161 |
| 3 | 119.7511 | 99.44330494 | 7.045 |
| 4 | 178.9902 | 63.40127166 | 2.8758 |
| 5 | 216.286 | 51.62196471 | 31.7121 |
| 6 | 244.7898 | 45.2034287 | 2.3669 |
| 7 | 272.6776 | 40.30082405 | 11.5537 |
| 8 | 293.7979 | 37.24187098 | 9.3093 |
| 9 | 313.6929 | 34.75678006 | 12.6033 |
| 10 | 323.0219 | 33.70224558 | 4.0873 |
| 11 | 355.5679 | 30.47637405 | 55.3009 |
| 12 | 381.0274 | 28.35339574 | 4.7475 |
| 13 | 413.5937 | 26.03367213 | 0.4502 |
| 14 | 453.4567 | 23.66383821 | 2.7012 |
| 15 | 464.7244 | 23.07023026 | 4.3244 |
| 16 | 465.2336 | 23.04410699 | 4.7418 |
| 17 | 487.3681 | 21.96304769 | 26.6401 |
| 18 | 493.5653 | 21.67831266 | 0.5839 |
| 19 | 515.983 | 20.70720939 | 8.2532 |
| 20 | 521.8814 | 20.46598799 | 1.553 |
| 21 | 548.4529 | 19.44552832 | 31.6512 |
| 22 | 576.7432 | 18.46527056 | 7.4546 |
| 23 | 590.0593 | 18.03728299 | 11.9956 |
| 24 | 600.1998 | 17.72443724 | 11.5497 |
| 25 | 624.3929 | 17.02014245 | 14.8927 |
| 26 | 666.6689 | 15.91506478 | 4.5386 |
| 27 | 670.1073 | 15.83146345 | 7.2329 |
| 28 | 702.3886 | 15.08739375 | 58.6262 |
| 29 | 713.3862 | 14.84962491 | 3.1284 |
| 30 | 723.2229 | 14.64321588 | 45.2814 |
| 31 | 739.6208 | 14.31159636 | 32.9956 |
| 32 | 750.7363 | 14.09521684 | 22.9111 |
| 33 | 776.3995 | 13.61979359 | 45.2281 |
| 34 | 795.1363 | 13.29245408 | 4.8162 |
| 35 | 816.798 | 12.93309592 | 5.1833 |
| 36 | 828.3955 | 12.74857072 | 34.0733 |
| 37 | 844.6483 | 12.49866249 | 6.3445 |
| 38 | 848.6988 | 12.43789847 | 65.6385 |
| 39 | 869.2367 | 12.13867098 | 22.3062 |
| 40 | 878.9062 | 12.00272006 | 12.7268 |
| 41 | 891.7028 | 11.82741621 | 35.1937 |
| 42 | 906.1996 | 11.63490667 | 3.5124 |
| 43 | 911.5578 | 11.56532944 | 15.6276 |
| 44 | 930.1559 | 11.33015519 | 149.9376 |
| 45 | 952.0285 | 11.06552726 | 14.8784 |
| 46 | 984.7246 | 10.69222263 | 24.2422 |
| 47 | 1017.9689 | 10.33762702 | 2.3202 |
| 48 | 1027.0072 | 10.24525136 | 1.5699 |
| 49 | 1032.478 | 10.19013498 | 1.5654 |
| 50 | 1037.3051 | 10.14199378 | 12.1643 |
| 51 | 1043.4498 | 10.08136613 | 2.1446 |
| 52 | 1051.7175 | 10.00092621 | 0.7964 |
| 53 | 1122.4829 | 9.361577851 | 19.2405 |
| 54 | 1146.094 | 9.166064907 | 43.1209 |
| 55 | 1155.227 | 9.092611434 | 14.7908 |
| 56 | 1158.3551 | 9.067723158 | 24.0756 |
| 57 | 1166.6494 | 9.002385597 | 68.1472 |
| 58 | 1187.1883 | 8.844573961 | 5.1293 |
| 59 | 1209.0698 | 8.682421997 | 194.0887 |
| 60 | 1251.9215 | 8.381497276 | 64.2387 |
| 61 | 1269.6072 | 8.263295377 | 5.2541 |
| 62 | 1272.0494 | 8.247234467 | 149.5957 |
| 63 | 1284.4341 | 8.166739294 | 32.1412 |
| 64 | 1291.8045 | 8.119576401 | 4.0088 |
| 65 | 1328.3917 | 7.893295336 | 14.9483 |
| 66 | 1333.4459 | 7.863024271 | 61.0385 |
| 67 | 1351.0275 | 7.75950779 | 46.404 |
| 68 | 1385.3721 | 7.564959876 | 248.6709 |
| 69 | 1401.8814 | 7.474871998 | 101.2585 |
| 70 | 1412.3529 | 7.418834894 | 16.7663 |
| 71 | 1425.942 | 7.347354864 | 118.1941 |
| 72 | 1436.3533 | 7.293515465 | 6.2199 |
| 73 | 1439.445 | 7.277679138 | 2.3251 |
| 74 | 1461.9206 | 7.164589497 | 93.5924 |
| 75 | 1476.5991 | 7.092610398 | 161.1752 |
| 76 | 1485.5499 | 7.04942381 | 29.6213 |
| 77 | 1513.8494 | 6.916276606 | 55.3083 |
| 78 | 1523.87 | 6.870328155 | 28.4212 |
| 79 | 1536.3292 | 6.814042235 | 23.9083 |
| 80 | 1538.8101 | 6.802944399 | 9.6221 |
| 81 | 1572.4038 | 6.656151425 | 5.4709 |
| 82 | 1595.2885 | 6.559728217 | 11.851 |
| 83 | 1602.4134 | 6.53027563 | 47.8681 |
| 84 | 1622.7492 | 6.447648731 | 324.2537 |
| 85 | 1635.575 | 6.396602636 | 16.9662 |
| 86 | 3224.333 | 3.229472064 | 0.4359 |
| 87 | 3225.9754 | 3.227819923 | 4.2773 |
| 88 | 3226.2433 | 3.227550594 | 0.5089 |
| 89 | 3231.1957 | 3.222579869 | 2.4154 |
| 90 | 3238.4155 | 3.215360722 | 4.9036 |
| 91 | 3240.962 | 3.212822162 | 3.8367 |
| 92 | 3245.6243 | 3.208184773 | 6.3155 |
| 93 | 3251.6433 | 3.202217693 | 17.5777 |
| 94 | 3262.1875 | 3.191817775 | 8.4583 |
| 95 | 3268.8179 | 3.185312627 | 45.5112 |
| 96 | 3281.531 | 3.172913558 | 34.7141 |

Ndft : DFT calculated harmonic frequency ($cm^{-1}$)

λ(μm) : Corrected wavelength (μm), scale factor=0.965, red shift= $-15 cm^{-1}$

ε(km/mol) : IR absorption coefficient (integral intensity)

Submitted to arXiv.org : Feb. 6, 2015 by Norio Ota